\documentclass[amsmath,amssymb,superscriptaddress,aps,pra,10pt]{revtex4-2}

\usepackage[utf8]{inputenc}
\usepackage{graphicx}
\graphicspath{{images/}}
\usepackage{geometry}
\usepackage[export]{adjustbox}

\usepackage[dvipsnames]{xcolor}
\usepackage{soul}

\usepackage{textcomp}
\usepackage{multirow}
\usepackage{array}
\usepackage{upgreek}
\usepackage[english]{babel}
\usepackage{longtable}
\usepackage[colorlinks=true, linkcolor=blue, citecolor=blue, urlcolor=blue]{hyperref}
\usepackage{needspace}

\usepackage[caption=false]{subfig}
\usepackage[amssymb]{SIunits}

\begin{document}

\title{The effects of salinity and inclination on the morphology of melting ice}

\author{Tomás J.~Ferreyra Hauchar}
\affiliation{Physics of Fluids department, Max Planck UT Center for Complex Fluid Dynamics, Faculty of Science and Technology, J.M. Burgers Centre for Fluid Dynamics, University of Twente, P.O. Box 217, 7500 AE Enschede, Netherlands.}
\author{Detlef Lohse}
\affiliation{Physics of Fluids department, Max Planck UT Center for Complex Fluid Dynamics, Faculty of Science and Technology, J.M. Burgers Centre for Fluid Dynamics, University of Twente, P.O. Box 217, 7500 AE Enschede, Netherlands.}
\affiliation{Max Planck Institute for Dynamics and Self-Organization, Am Fa\ss berg 17, 37077 G\"{o}ttingen, Germany}
\author{Sander G.~Huisman}
\thanks{ Contact author: s.g.huisman@utwente.nl} 
\affiliation{Physics of Fluids department, Max Planck UT Center for Complex Fluid Dynamics, Faculty of Science and Technology, J.M. Burgers Centre for Fluid Dynamics, University of Twente, P.O. Box 217, 7500 AE Enschede, Netherlands.}

\date{\today}

\begin{abstract}
The salinity of water and the slope of ice significantly influence the melt rate and surface morphology of ice, both highly relevant in the context of glacier and iceberg melting in oceanic environments. In this study, we conducted experiments on vertical and sloped ice blocks melting in quiescent saline water. Through the use of fringe projection profilometry, we measured the morphology of the ice's front face. In particular, we combine the spatio-temporal phase shifting and orthogonal sampling moire methods. The far field salinity in the experiments ranged from $0$ g/kg to $35$ g/kg, and angles were between \unit{-18}{\degree} and \unit{50}{\degree}. The ice blocks were $L\times W \times H = \unit{12}{\centi\meter} \times \unit{23}{\centi\meter} \times \unit{32}{\centi\meter}$ and submerged in water of approximately \unit{20}{\celsius}, leading to $\textrm{Ra} = \mathcal{O}(10^7)$. We identified and classify five surface morphologies and regimes arising from the flow regimes imposed by salinity and inclination, namely scalloped, channelized, top-melting, bottom-melting, and incurved. The channelized morphology consists of vertical channels carved along the ice surface, whose development originates from a Rayleigh--Bénard type instability, and which are enhanced by bubbles released from the melting ice and rising along the interface. The scalloped regime is characterized by a rough dimpled pattern commonly referred to as scallops. We observe that increasing the salinity leads to scallops that are smaller, shallower, and more uniform in size. Additionally, a salinity dependence of the melt rate is found, showing a non-monotonic behavior, while the inclination angle shows little influence on the overall melt rate.
\end{abstract}

\maketitle

\section{Introduction} \label{sec:intro}

In recent decades, the loss of ice on Greenland and Antarctica has accelerated \cite{Rignot2011}, affecting ocean currents and driving sea level rise. Despite this, model predictions can be up to 100 times lower than observed melt rates \cite{Sutherland2019}. Recent observations on Thwaites Glacier \cite{Schmidt2023,Lopatka2023} show a strong dependence of melt rate on small-scale structures. In particular, a strong influence of the interfacial slope is observed, with angles lower than \unit{60}{\degree} with respect to gravity accounting for 27\% of the melt rate, even when they account for only 9\% of the ice surface \cite{Schmidt2023}. Moreover, images from beneath ice shelves and free-drifting icebergs show rough patterns on the surface of the ice \cite{Lawrence2023,Hobson2011,Schmidt2023}. As the climate continues to heat up, increased physical understanding of ice sheet-ocean interaction will be required to improve our predictions \cite{Malyarenko2020,Cenedese2023,Du2024}.

We simplify the naturally-occurring melting ice by focusing on the study of rectangular ice blocks submerged in quiescent saline water at room temperature (between \unit{18}{\celsius} and \unit{21}{\celsius}). Water temperatures near \unit{4}{\celsius} were deliberately avoided to prevent additional complications associated with the density anomaly of water. Both salinity and inclination of the ice blocks were varied, leading to distinct patterns forming on the surface of the ice. Dimpled patterns, similar to the ones seen in nature and usually referred to as scallops, have previously been observed in experiments and simulations of natural convection \cite{Josberger1981,xu2024,Weady2022}, forced convection \cite{Bushuk2019,Perissutti2024}, and dissolution \cite{DaviesWykes2018,Cohen2020}. For varying salinity and temperature the flow dynamics can roughly be categorized into thermally-driven, salinity-driven, and competing regimes \cite{xu2024}, which differ in the relative driving strength of the thermal and solutal boundary layers. The thermally-driven regime is characterized by a cold thermal boundary layer that descends along the wall. In the salinity-driven regime there is a cold upward flow driven by the lower density of fresh melt water. Finally, in the competing regime there is an inner solutal boundary layer flowing up the wall, which is embedded in a downward flowing thermal boundary layer. This effect is caused by the ratio between the thermal diffusivity $\kappa_T$ and mass diffusivity $\kappa_S$, expressed as the Lewis number $\textrm{Le} = \kappa_T / \kappa_S$ which is of $\mathcal{O}(100)$ for water. Because heat diffuses much faster than salt, the thermal boundary layer is thicker than the solutal one, with their thickness ratio scaling as $\delta_T / \delta_S \propto \sqrt{\textrm{Le}} \approx 10$, giving rise to the competing regime. Scallop patterns were only seen to develop within the competing regime. While Xu et al. \cite{xu2024} focused on flow regimes and their effects, Chaigne et al. \cite{Chaigne2023} proposed a geometrical approach, with a non-uniform normal ablation model to explain the pattern formation. By setting a spatially heterogeneous erosion rate over an initially flat plane, they show the emergence of a pattern consisting of cavities surrounded by sharp crests. Claudin et al. \cite{Claudin2017} studied the growth or decay of a sinusoidally perturbed dissolving surface, through a purely theoretical approach. They used a turbulent mixing description that accounts for the bed roughness. However, the studied problem differs from ours since it doesn't include thermal gradients inside the ice, and only considers two boundary layer (viscous and solutal) instead of three (viscous, solutal, and thermal). 

McConnochie and Kerr \cite{McConnochie2018} conducted experiments of sloped dissolving ice. They propose that the ablation velocity scales with $\cos(\theta)^{2/3}$ relative to the vertical case. However, when comparing this model to observed melt rates under Thwaites Glacier, it was found that this parametrization underestimates the melt rate \cite{Schmidt2023}. A buoyant plume model predicts a scaling of $\cos(\theta)^{-1}$ for near vertical walls, and $\cos(\theta)^{3/2}$ for high angles \cite{Magorrian2016}. Note that these regimes are relevant for ice heights of hundreds of meters, well outside the scope of lab experiments. Simulations of sloping ablating surfaces \cite{Mondal2019} support the $\cos(\theta)^{2/3}$ scaling for polar ocean conditions with salinities near 35 g/kg and temperatures lower than \unit{5}{\celsius}. 

In this work we investigate the effect of inclination and salinity on the melting rate of inclined ice blocks. We focus on salinities ranging from 0 to ocean conditions (35 g/kg) and temperatures between \unit{18}{\celsius} and \unit{21}{\celsius}.
Additionally, we use fringe projection profilometry (FPP) to measure the morphology formed on the surface of the ice. Although FPP has previously been used in the context of fluid dynamics \cite{Cobelli2009}, to our knowledge this has never been used for the study of pattern formation on dissolving or melting bodies. In § \ref{sec:setup} we describe the experimental set up, while in § \ref{sec:fpp} the details on the implementation of FPP can be found. In § \ref{sec:theory}, we provide a brief theoretical background on the dynamics of a melting ice face. Following that, § \ref{sec:results} presents the results of our experiments, including the effect of salinity and inclination on melt rate, as well as the different morphologies found. Finally, concluding remarks are provided in § \ref{sec:conclusion}.

\section{Experimental setup} \label{sec:setup}

\begin{figure}[t]
    \centering
    \includegraphics[width=0.7\linewidth]{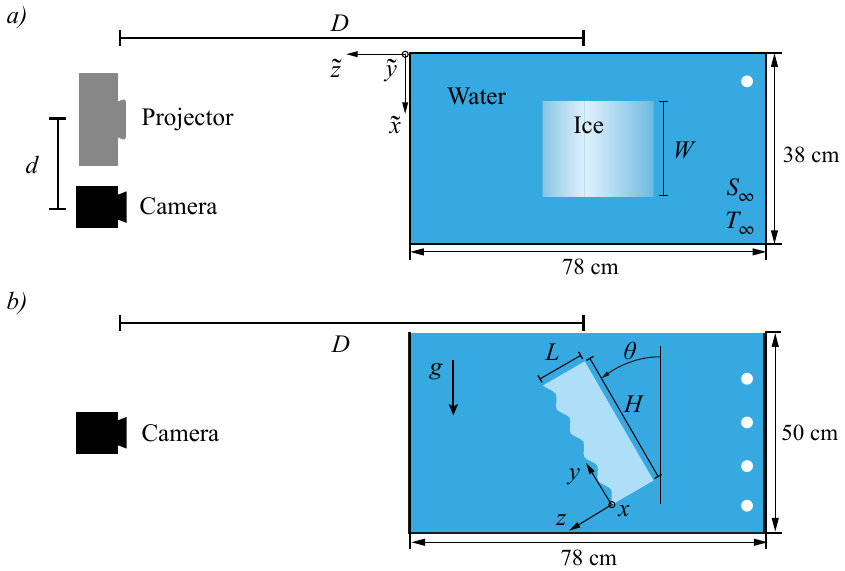}
    \caption{ Top $a)$ and side $b)$ views of the experimental setup. A rectangular ice block of dimensions $H\times W\times L$ is placed in the center of the glass tank. A projector and camera, with parallel optical axes, are at a distance $D$ from the ice block, while being a distance $d$ from each other. The ice block is inclined with an angle $\theta$, where the arrow denotes positive angles. The white dots indicate the locations of the temperature and salinity probes within the tank.} 
    \label{fig:setup}
\end{figure}

The experimental setup, depicted in Figure \ref{fig:setup}, consisted of rectangular ice blocks placed inside a water tank filled with quiescent water, a projector, and a camera.
The NaCl concentration, i.e. the salinity, of the water and the inclination angle of the ice blocks were systematically varied. The inner dimensions of the water tank were 50 cm in height, 38 cm in width, and 78 cm in length. The ambient water temperature $T_\infty$ ranged from \unit{18}{\celsius} to \unit{21}{\celsius}, and the salinity $S_\infty$ was varied between 0 g/kg and 35 g/kg. An ice block was placed centrally within the tank, and its inclination angle ($\theta$) with respect to the vertical was varied from \unit{-15}{\degree} to \unit{50}{\degree}. Ambient temperature and salinity were measured using StarOddi DST CT probes, which provided temperature accuracy of \unit{\pm 0.1}{\kelvin} and salinity accuracy of ±1 g/kg. These probes were positioned at distances of 6.5 cm, 16.5 cm, 27.5 cm, and 38.5 cm from the tank's bottom.

The imaging system consisted of a Photron Fastcam Mini AX200 camera and an Optoma UHZ65LV projector. The projector was used for FPP, which is described in the following section (§ \ref{sec:fpp}). Images were projected at 15 frames per second (fps) and 4K resolution, and recorded at 60 fps, resulting in multiple captures of the same projected image. The recorded image size was 1024 px $\times$ 1024 px and the field of view was around 39 cm $\times$ 39 cm, resulting in a resolution of about 0.4 mm/px. A pixel-wise average of the repeated images was computed to reduce noise. The camera and projector were aligned such that their optical axes were parallel. They were positioned $d=60$ cm apart, with a distance of approximately $D=200$ cm from the ice block.

The implementation of FPP requires the ice blocks to be opaque. To achieve this, $\mathcal{O}$(1 ml) of Edding 90 white marker ink was mixed with the water ($\approx 100$ ppm) used to create the ice blocks. The water–ink mixture was cooled to near \unit{0}{\celsius} in an 8 mm thick aluminum mold. The mold was then placed inside a larger vessel into which liquid nitrogen was poured to induce rapid freezing of the mixture. The rapid freezing process ensured uniform freezing of the dye, that also resulted in the formation of tiny bubbles ($<1$ mm) within the ice, further enhancing opacity. The liquid nitrogen was insufficient to freeze the interior of the blocks completely, so the blocks were left to fully freeze at \unit{-16}{\celsius} for at least 10 hours. After unmolding, the ice blocks were stored at either \unit{-5}{\celsius} or \unit{-16}{\celsius} for 24 hours to thermally equilibrate them before the experiment. On the other hand, a few experiments were also performed using mostly clear ice. As expected, FPP could not be applied to these experiments, and they were therefore used only for qualitative comparison. To produce the mostly clear ice blocks, water was frozen in the same aluminum molds at \unit{-5}{\celsius}. During freezing, the forming ice layer at the top of the mold was manually broken every few hours to allow entrapped air to escape and promote freezing from the bottom upward. During non-working hours, the remaining unfrozen water was removed to prevent freezing from the top.

Three different sizes of ice blocks were used, grouped into three sets. The first set consisted of ice blocks with a height, a width, and a length of $H \times W \times L = 32~\mathrm{cm} \times 17~\mathrm{cm} \times 10~\mathrm{cm}$. The second set had ice blocks with dimensions of $H \times W \times L = 32~\mathrm{cm} \times 23~\mathrm{cm} \times 12~\mathrm{cm}$, while the third set had dimensions of $H \times W \times L = 23~\mathrm{cm} \times 15~\mathrm{cm} \times 12~\mathrm{cm}$. Further details on each experiment can be found in Appendix \ref{sec:table_exp}.

\section{Morphology measurement technique} \label{sec:fpp}

To capture the temporal evolution of ice surface morphology, we employed fringe projection profilometry (FPP), which reconstructs the height profile from the deformation of a projected pattern. The surface height $h$ is related to the phase difference $\Delta\varphi = \varphi - \varphi_0$ between the deformed and reference patterns. Following the formulation by Takeda et al. \cite{Takeda1983} and refined by Maurel et al. \cite{Maurel2009} we have:
\begin{align}
    h(\tilde{x}',\tilde{y}') &= \frac{\tilde{D} \Delta\varphi(\tilde{x},\tilde{y})}{\Delta\varphi(\tilde{x},\tilde{y}) - \omega \tilde{d}} \label{eq:height_profile} \\
    \tilde{x}' &= \tilde{x} - \frac{h(\tilde{x}',\tilde{y}')}{\tilde{D}} \tilde{x} \\
    \tilde{y}' &= \tilde{y} - \frac{h(\tilde{x}',\tilde{y}')}{\tilde{D}} \tilde{y}
\end{align}
where $\omega$ is the spatial frequency of the fringe pattern, and $\tilde{x}$ and $\tilde{y}$ are the coordinates on the reference plane. $\tilde{D}$ denotes the distance from the projector and camera to the object surface, incorporating the changes in media effects. Similarly, $\tilde{d}$ represents distance between the projector and the camera optical axis, with change in media incorporated. $\tilde{x}'$ and $\tilde{y}'$ are corrections to where the height profile is measured. A 3D-printed calibration target of known geometry was used to determine the optimal values of $\tilde{D}$ and $\omega\tilde{d}$ for each experiment.

A variety of phase-retrieval methods have been developed, many of which can be viewed as extensions or variations of phase-shifting profilometry (PSP). We therefore briefly review the PSP framework before describing the phase-retrieval method employed in the present work. In PSP, a sequence of $N$ sinusoidal fringe patterns is projected onto the object surface, each described by \cite{Zuo2018}
\begin{align}
    I_n^p(\tilde{x},\tilde{y}) = a^p + b^p \cos(\omega \tilde{x} + 2\pi n/N) = a^p + b^p \cos(\varphi_0 + 2\pi n/N) ,
\end{align} 
where $n$ denotes the phase-shift index ($1\leq n \leq N$), and $a^p$ and $b^p$ represent the mean and modulation intensities of the projected pattern, respectively. After reflection from the object surface, the fringe pattern is recorded by the camera as
\begin{align}
    I_n = A(\tilde{x},\tilde{y}) + B(\tilde{x},\tilde{y}) \, \cos\left(\varphi(\tilde{x},\tilde{y}) + 2\pi n/N\right),
\end{align}
where $A(\tilde{x},\tilde{y})$ and $B(\tilde{x},\tilde{y})$ denote the background and amplitude intensities, respectively. The phase field can then be recovered from
\begin{align}
    \varphi = \arg \left[ \sum_{n=1}^{N} I_n(\tilde{x},\tilde{y}) \,\, e^{-i \frac{2\pi n}{N}} \right].
    \label{eq:varphi}
\end{align}
The reference phase $\varphi_0$ can be obtained analogously using a flat reference surface, yielding the phase difference $\Delta\varphi=\varphi-\varphi_0$.

In the present work, the phase difference was determined using the spatiotemporal phase-shifting method (ST-PSM) \cite{Ri2019}. ST-PSM extends the conventional PSP framework by incorporating sampling moiré (SM) \cite{Ri2010}, improving robustness in low-contrast imaging conditions. In the SM approach, only one fringe image is recorded, which is then subsampled with a pitch $P$ smaller than the fringe period, generating $P$ phase-shifted images (see Ri et al. \cite{Ri2010} for a detailed explanation on subsampling). Applying this procedure to each of the $N$ PSP images yields a set of $P\times N$ images, $I_{p,n}(\tilde{x},\tilde{y})$, from which the phase difference is calculated as
\begin{align} 
    \Delta\varphi = \arg \left[ \sum_{p=1}^{P}\sum_{n=1}^{N} I_{p,n}(\tilde{x},\tilde{y}) \,\, e^{-i \frac{2\pi p}{P}} e^{-i \frac{2\pi n}{N}} \right].
\end{align} 

However, it is known that phase-shifting methods can generate fringe like artifacts in the phase calculation \cite{Lu2017,Xing2019}, and ST-PSM isn't enough to avoid them. By combining ST-PSM with orthogonal sampling moiré (OSM) \cite{Chen2022} we were able to mitigate these artifacts. We will call the combination of these two techniques orthogonal spatiotemporal phase-shifting method (OST-PSM) from now on.

For OST-PSM, the projected patterns are $I_{m,n}^{\text{p}} = a^p + b^p \cos(\omega \tilde{x} + 2\pi m/N)\cos(\omega \tilde{y} + 2\pi n/N)$, where $1 \leq m,n \leq N$. The recorded image is subsequently subsampled in both spatial directions with sampling pitch $P$ to obtain the moiré intensity $I^{\text{moi}}_{m,n,p,q}$, where $1 \leq p,q \leq P$ index the subsampling in the $\tilde{x}$ and $\tilde{y}$  directions, respectively. Subsampling enhances the accuracy of the height measurements at the expense of spatial resolution. Following ST-PSM and OSM, the phase $\varphi$ for OST-PSM can be calculated from the moire intensity image as 
\begin{align} 
    \varphi(\tilde{x},\tilde{y}) =  \arg \left[ \sum_{p=1}^{P}\sum_{q=1}^{P} \sum_{m=1}^{N} \sum_{n=1}^{N} I^{\text{moi}}_{m,n,p,q}(\tilde{x},\tilde{y}) \,\, e^{-i \frac{2\pi m}{N}} e^{-i \frac{2\pi p}{P}} \right].
\end{align} 
where we chose $N=5$, totalling 25 projected images, and $P=11$ as the amount of subsampling steps. The same procedure was applied to obtain the reference phase $\varphi_0$, with the projected pattern reflected from a flat plate. 

After reconstructing the surface profiles for each experiment, the profiles were rotated to align with the initial imposed orientation, coordinates $x,y,z$. A Gaussian filter with standard deviation $\sigma = 2$ was applied to suppress noise and remove minor artifacts. Finally, the data points were interpolated onto a rectangular grid to facilitate subsequent data analysis.

\section{Background} \label{sec:theory}

\begin{figure}[t]
    \centering
    \includegraphics[width=1.0\linewidth]{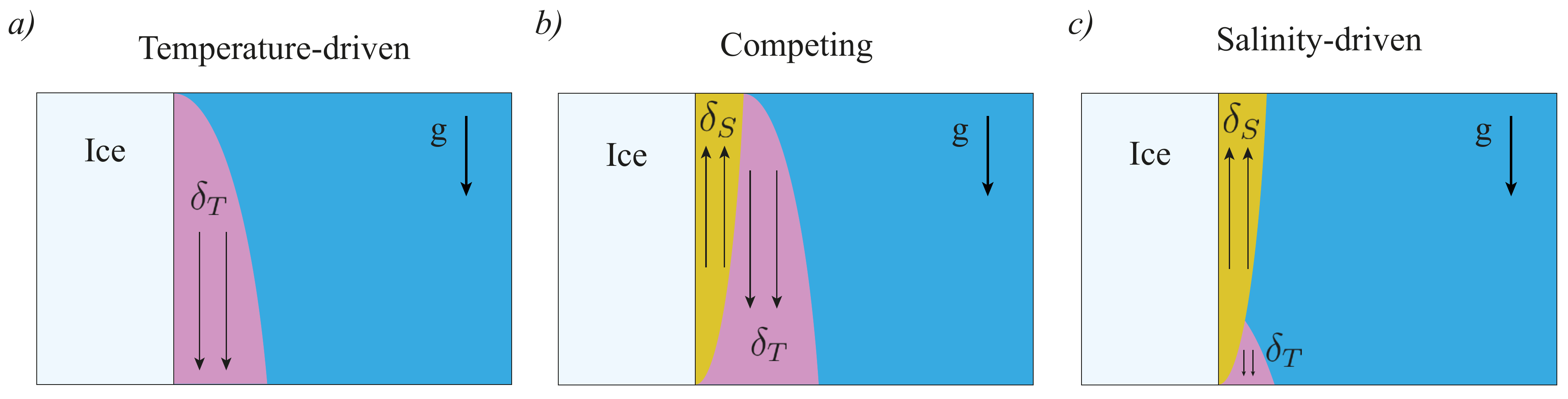}
    \caption{Side view sketch of the boundary layers for the three different regimes. The boundary layer thicknesses are not shown to scale. $a)$ In the temperature-driven regime ($R_\rho < 1$) the thermal boundary layer is denser than the ambient water, therefore it flows down. $b)$ A inner solutal boundary layer forms in the competing regime ($R_\rho \approx 2$). This layer is less dense than the outer thermal boundary layer, creating a bidirectional flow. $c)$ In the salinity driven regime ($R_\rho \gg 1$) the driving of the solutal boundary layer is much stronger than the thermal boundary layer, except near the leading edge at the bottom where the solutal boundary layer is forming.}
    \label{fig:sketches}
\end{figure}

The dynamics of ice melting in saline water is controlled by 3 fields: temperature, salinity, and velocity. In the absence of external forcing, fluid motion arises solely from buoyancy differences, which originate from variations in temperature and salinity. Both temperature and salinity vary sharply within their respective boundary layers, where conditions transition from the interface values $T_m$ and $S_m$ to the ambient conditions $T_\infty$ and $S_\infty$. For the range of ambient temperatures investigated in this work (\unit{\approx 20}{\celsius}), the interfacial temperature and salinity can be approximated as $T_m = \unit{0}{\celsius}$ and $S_m = 0$ g/kg \cite{Wells2011}.

The melting of ice blocks in water falls into the wider category of Stefan problems \cite{Stefan1891,Rubinstein1971}, which describes the interface evolution of a two phase system undergoing phase transition. From the heat flux balance at the ice-water interface we get
\begin{equation}
    u \rho_{\text{ice}} \mathcal{L} = k_{\text{ice}} \frac{\partial T}{\partial z}\bigg|_{-} - k_{w} \frac{\partial T}{\partial z}\bigg|_{+},
    \label{eq:stefan}
\end{equation}
where the suffixes $_{\text{ice}}$ and $_w$ indicate whether it is ice or water, $u$ is the melting rate (with units of m/s; it is the velocity with which the interface recedes), $\rho$ is the density, $\mathcal{L}$ is the latent heat of fusion, and $k$ the thermal conductivity. The equation states that the heat flux due to latent heat is equal to conduction of heat at the water ($+$) and ice ($-$) sides of the interface. The heat flux at the interface is typically made dimensionless using the conductive heat transfer to form the Nusselt number
\begin{equation}
    \textrm{Nu} = \frac{\text{heat flux}}{\text{conductive heat flux}} = \frac{u \, \rho_{\textrm{ice}} \, \mathcal{L}\, H}{k\, (T_\infty - T_m) } = \frac{\rho_{\textrm{ice}}}{\rho_w} \frac{1}{\textrm{Ste}} \frac{H}{ \kappa_T} u
\end{equation}
to describe the dimensionless heat flux at the interface, with $H$ the height of the ice block, $\textrm{Ste} = c_p \Delta T / \mathcal{L} $ the Stefan number, $c_p$ the heat capacity of water, $\Delta T = T_\infty - T_m$, and  $\kappa_T$ the thermal diffusivity of water. The mean melting rate $u$ can be calculated from the interface position $h(x,y,t)$ as follows 
\begin{equation}
    u = \frac{1}{\tau} \int_0^{\tau} \left[ \frac{1}{A(t)} \iint -\frac{\partial h(x,y,t)}{\partial t} \, dA(t) \right] dt.
\end{equation}

Fluid motion is driven by buoyancy contrasts that arise as meltwater of differing temperature and salinity mixes with the ambient fluid. For sloping surfaces, the strength of this buoyancy forcing is characterised by a Rayleigh number
\begin{equation}
        \textrm{Ra} = \frac{g \cos(\theta) \, H^3}{ \kappa_T \nu} \frac{| \rho(T_m,S_m) - \rho(T_\infty,S_\infty)|}{\rho(T_\infty,S_\infty)},
\end{equation}
with $\nu$ the kinematic viscosity of water, and $g$ the gravitational acceleration.
Because both temperature and salinity contribute to density, it is useful to distinguish their respective strength using the density ratio
\begin{equation}
    R_{\rho} = \frac{\Delta\rho_S}{\Delta\rho_T } = \frac{|\rho(T_\infty,S_\infty) - \rho(T_\infty,S_m)|}{|\rho(T_\infty,S_\infty) - \rho(T_m,S_\infty)|}.
\end{equation}
It is worth noting that we chose to calculate density differences against the ambient conditions. However, alternative formulations are possible, for example using thermal expansion and haline contraction coefficients or evaluating density differences relative to interface conditions. We adopt ambient conditions as the reference state since the density profile is nonlinear in the temperature–salinity ranges considered, and because ambient-based differences provide a more representative measure of the buoyancy forcing that drives the flow.

The value of $R_\rho$ defines three distinct flow regimes. For $R_\rho \ll 1$ we are in the temperature driven regime, for $R_\rho \gg 1$ we have the salinity driven regime, and finally the competing regime for $R_\rho \approx 1$. Figure \ref{fig:sketches} shows a qualitative sketch of the thermal and solutal boundary layers in these regimes.  In the temperature-driven regime (low salinity, Figure~\ref{fig:sketches}$a$), melt water is denser than the ambient water. That leads to downward flowing thermal boundary layer that thickens with depth. For the case there is not heat diffusion in the ice, according to Eq.~(\ref{eq:stefan}), the local melt rate is proportional to the temperature gradient at the ice--water interface. Approximating this gradient as $\partial T/\partial z \approx (T_\infty-T_m)/\delta_T$, where $\delta_T$ is the thermal boundary layer thickness, it follows that the melt rate is inversely proportional to the thermal boundary-layer thickness. Since the thermal boundary layer becomes progressively thicker with depth, the result is a faster melting at the top of the ice surface compared to the bottom \cite{xu2024}. For ice blocks with a positive inclination, the denser melt water layer would lie on top of the ambient water, leading to a Rayleigh--Bénard instability \cite{Cohen2020}.  

For $R_\rho \approx 2$ a thin solutal boundary layer of fresh meltwater forms within a thicker thermal boundary layer of cold, salty water, Figure~\ref{fig:sketches}$b$. The ratio between thermal and mass diffusivities, known as the Lewis number ($\textrm{Le}=\kappa_T/\kappa_S$), dictates the relation between boundary layers thicknesses. For water solution of NaCl at \unit{20}{\celsius}, the Lewis number is $\mathcal{O}(100)$, resulting in a thermal boundary layer 10 times thicker than the solutal one. Consequently, the solutal boundary layer flows upward, while the thermal flows downward, giving rise to a bidirectional flow near the interface.

Finally, in the salinity-driven regime ($R_\rho \gg 1$, Figure~\ref{fig:sketches}$c$) buoyancy forcing is dominated by salinity, and the strong upward solutal plume is capable of dragging a thin layer of the thermal boundary layer upward with it. Near the leading edge at the bottom, however, the solutal layer is not yet strong enough, allowing the thermal boundary layer to flow downward. Due to the diverging boundary layers, this bifurcation point advects warm ambient water towards the ice--water interface, leading to a faster local melt rate there \cite{Josberger1981}.

\section{Results} \label{sec:results}

\subsection{Morphology and heat transfer}

\begin{figure}
    \centering
    \includegraphics[width=0.9\linewidth]{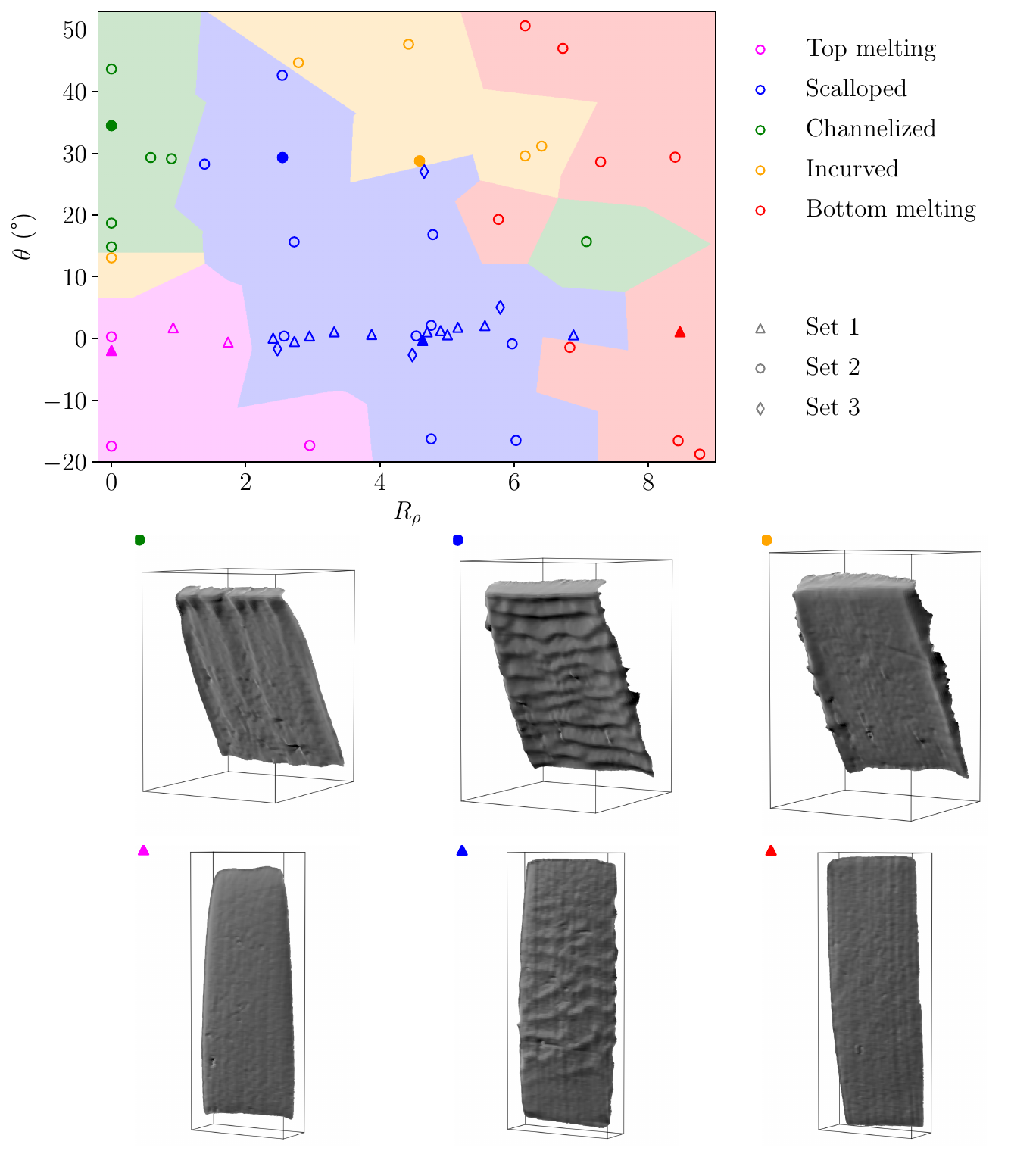}
    \caption{$R_\rho - \theta$ parameter space of the performed experiments, including the observed morphology at the end of each experiment. Background colors indicate the morphology at the nearest experimental data point (Voronoï construction), providing a visual guide to the distribution of morphologies in phase space. For the Voronoï construction, both axes are normalized by their respective ranges so that distances in the parameter space are treated equally. 3D renders of the different surface morphologies are shown on the bottom, see the supplemental material at \cite{supplement} for a movie of the renders. For reference, the bounding box height is 28 cm. Figures \ref{fig:profile_salinity} and \ref{fig:profile_inclined} show quantitative profiles for the experiments marked with filled triangles and disks, respectively.  }
    \label{fig:morpho}
\end{figure}

\begin{figure}
    \centering
    \includegraphics[width=1.0\linewidth]{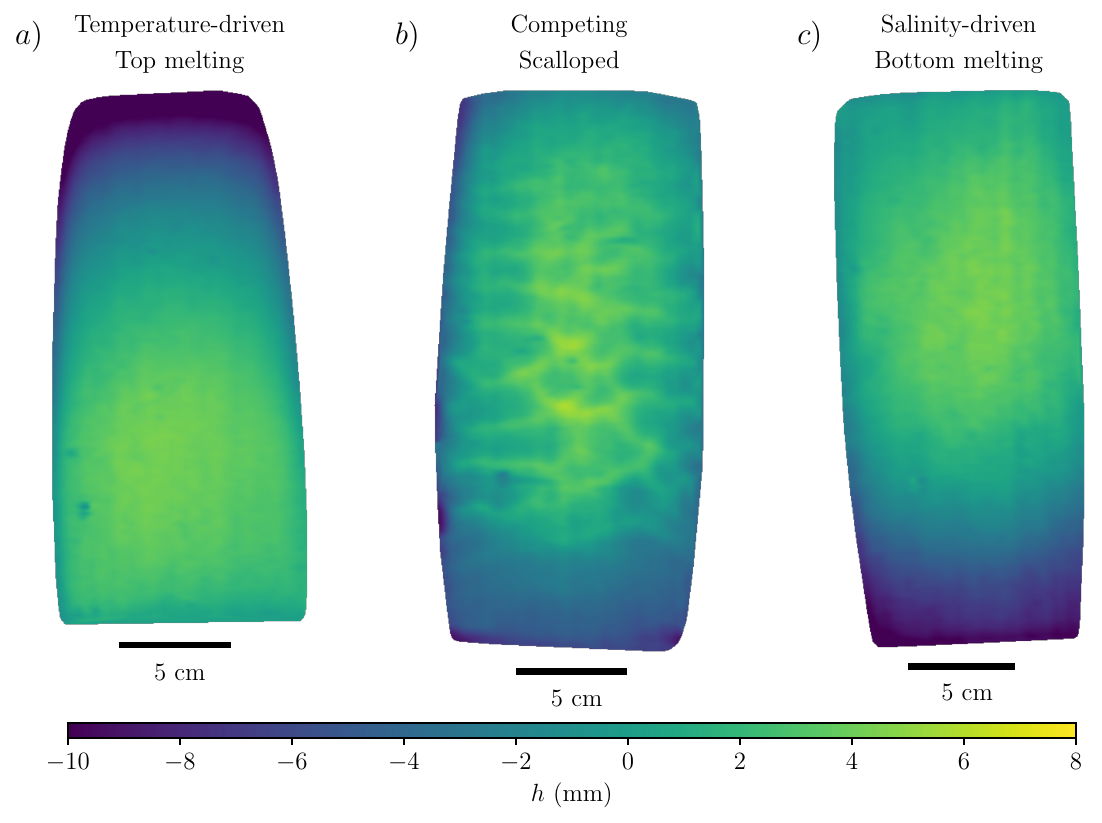}
    \caption{ Vertical ice faces: height profiles at 30 minutes of starting the experiment for three different flow regimes: $a)$ $S=0.0$ g/kg, $T_w = \unit{21.0}{\celsius}$, $\theta = \unit{-2.0}{\degree}$, Ra $= 4.1 \times 10^6$, and $R_\rho=0.0$; $b)$ $S=14.8$ g/kg, $T_w = \unit{20.0}{\celsius}$, $\theta = \unit{-0.3}{\degree}$, Ra $= 2.1 \times 10^7$, and $R_\rho=4.6$; $c)$ $S=27.4$ g/kg, $T_w = \unit{17.8}{\celsius}$, $\theta = \unit{1.1}{\degree}$, Ra $= 8.5 \times 10^7$, and $R_\rho=8.5$. }
    \label{fig:profile_salinity}
\end{figure}

\begin{figure}
    \centering
    \includegraphics[width=1.0\linewidth]{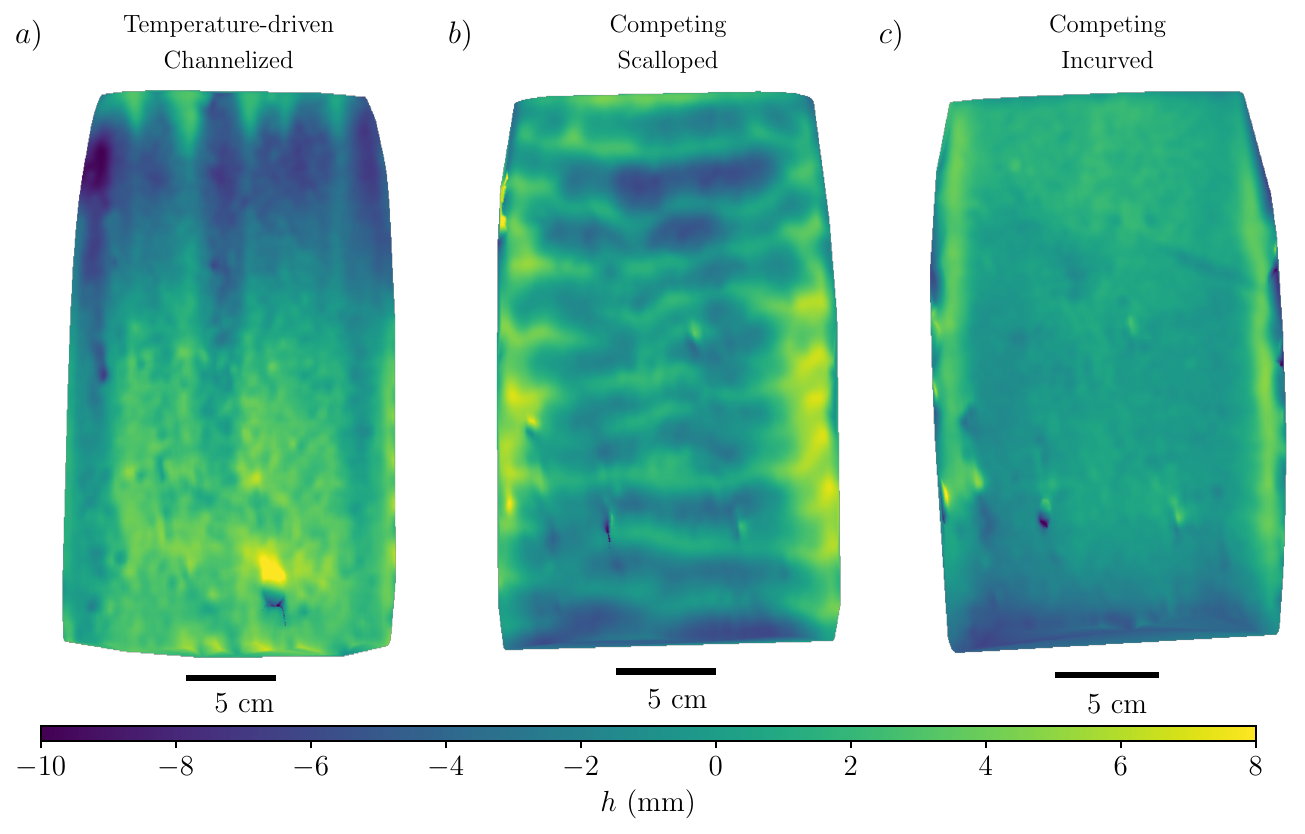}
    \caption{Tilted ice faces: height profiles at 30 minutes for two different flow regimes and three different morphologies: $a)$ $S=0.0$ g/kg, $T_w = \unit{19.0}{\celsius}$, $\theta = \unit{34.5}{\degree}$, Ra $= 2.6 \times 10^6$, and $R_\rho=0.0$; $b)$ $S=7.0$ g/kg, $T_w = \unit{20.3}{\celsius}$, $\theta = \unit{29.3}{\degree}$, Ra $= 7.0 \times 10^6$, and $R_\rho=2.6$; $c)$ $S=13.3$ g/kg, $T_w = \unit{19.4}{\celsius}$, $\theta = \unit{28.8}{\degree}$, Ra $= 1.7 \times 10^7$, and $R_\rho=4.6$. }
    \label{fig:profile_inclined}
\end{figure}

We first examine the range of surface morphologies that develop during the melting of ice blocks under varying experimental conditions. Based on the observed surface features, five distinct regimes were identified in the $S - \theta$ parameter space, for $\text{Ra}=\mathcal{O}(10^7)$ (Figure~\ref{fig:morpho}). Blue data points correspond to experiments where scallops form on the ice surface (Figures~\ref{fig:profile_salinity}$b$ and \ref{fig:profile_inclined}$b$), while green points indicate the formation of vertical channels (Figure~\ref{fig:profile_inclined}$a$). Magenta and red points correspond to cases dominated by top and bottom melting, respectively (Figures~\ref{fig:profile_salinity}$a$ and \ref{fig:profile_salinity}$c$). Experiments shown in orange exhibit a concave, or incurved, surface morphology in which the lateral edges melt more slowly than the center (Figure~\ref{fig:profile_inclined}$c$). These regimes are hereafter referred to as scalloped, channelized, top-melting, bottom-melting, and incurved. All experiments with $\theta > \unit{10}{\degree}$ show an incurved morphology. We chose to classify the morphology as incurved only if it didn't display any characteristic of the other morphologies, as it is likely caused by edge effects. A more detailed explanation on the morphology classification can be found in Appendix \ref{sec:morphologies}. The background colors in Figure~\ref{fig:morpho} serve as a visual guide to the regions of phase space in which each morphology occurs. The colors are assigned based on the morphology occurring at the nearest experimental data point (Voronoï construction). It is intended as an aid to visualization, and should not be interpreted as exact boundaries between morphological regimes. Three-dimensional renders of representative ice surfaces are shown at the bottom of Figure~\ref{fig:morpho}, and a video of the surface evolution over time for the shown experiments is available in the supplemental material at \cite{supplement}.

In Figures \ref{fig:profile_salinity} and \ref{fig:profile_inclined}, the experiments are categorized into three distinct regimes, depending on whether temperature, salinity, or neither serves as the dominant driving mechanism. In the temperature and salinity driven regimes we observe top and bottom melting morphologies, respectively. This is to be expected from the flow regimes described in Figure \ref{fig:sketches}. Top melting is observed for $R_\rho < 2$ in the case of vertical ice blocks ($\theta \approx \unit{0}{\degree}$), and $R_\rho \lesssim 3$ for ice blocks inclined with a negative slope. This slightly differs from $R_\rho < 1$ for the temperature driven regime previously suggested, however this can be attributed to the way we calculate $R_\rho$. On the other side, bottom melting occurs for $R_\rho \gtrsim 6$, although it is not a clearly defined transition. Moreover, we do not observe the transition point near the bottom of the ice, as expected from the flow regime (Figure \ref{fig:sketches}$c$). In both regimes, the transport of meltwater leads to a gradual stratification of the ambient fluid, with colder water accumulating near the bottom and top of the tank in the temperature- and salinity-driven regimes, respectively. While this stratification may further enhance the observed melting patterns at later times, preferential melting is already evident within the first few minutes of the experiments, indicating that the top- and bottom-melting morphologies are primarily governed by the local buoyancy-driven flow adjacent to the ice surface.

In the temperature-driven melting regime, we observe that for inclined ice blocks, distinct surface channels develop when the inclination angle exceeds approximately \unit{15}{\degree}. An exception occurs at higher salinities ($R_\rho \approx 7$), where the observed channels for this experiment differs from those of the other cases. This case presents channels that are comparatively shallower, while the channels edges are not narrow like the ones at low salinities, see Figure~\ref{fig:profile_inclined}$a$. Additionally, the characteristic feature of bottom melting is present, in accordance with the surrounding experiments. For these reasons, this case is excluded from the subsequent analysis of the channelized morphology.

\begin{figure}
    \centering
    \includegraphics[width=1.\linewidth]{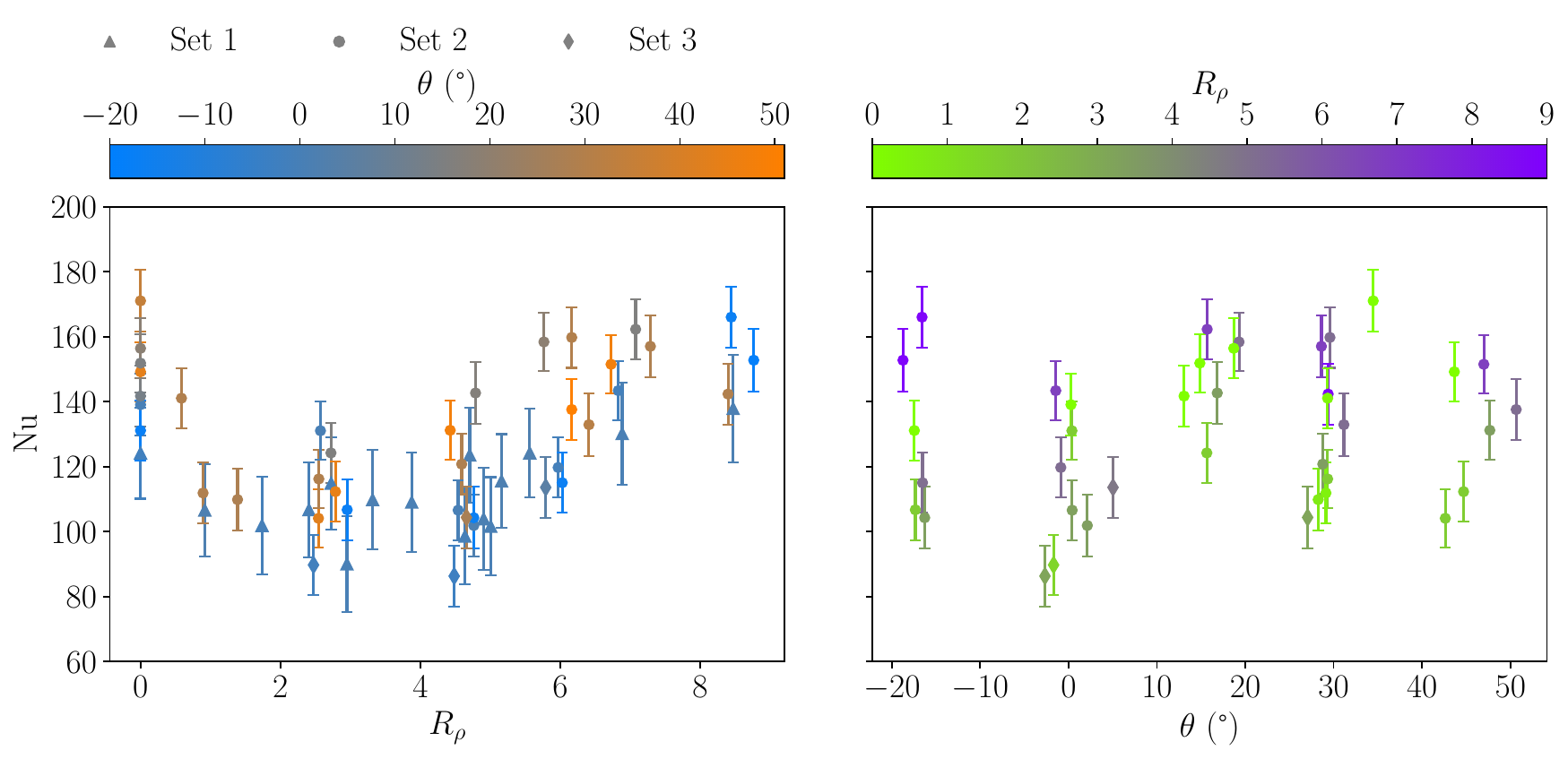}
    \caption{Nusselt number as function of the density ratio $R_\rho$ and inclination $\theta$. Set 1 was composed of ice blocks with a starting height, a width, and a length of $H \times W \times L = 32~\mathrm{cm} \times 17~\mathrm{cm} \times 10~\mathrm{cm}$. Set 2 had ice blocks with dimensions of $H \times W \times L = 32~\mathrm{cm} \times 23~\mathrm{cm} \times 12~\mathrm{cm}$, while set 3 had dimensions of $H \times W \times L = 23~\mathrm{cm} \times 15~\mathrm{cm} \times 12~\mathrm{cm}$.} 
    \label{fig:vertical_mr}
\end{figure}

These channels are similar to those seen in icebergs \cite{Hobson2011} and ice shelves \cite{Vreugdenhil2026}, which are also referred to as runnels \cite{Vreugdenhil2026}, and have also been reported in forced convection simulations \cite{Couston2021}. In the context of natural convection, similar structures have also been previously reported for dissolution experiments \cite{Cohen2020}, but with a key difference: in our experiments, the channels deepen preferentially near the upper portion of the ice while remaining absent near the lower edge. This asymmetry was not noted in the earlier study, although an entry length near the top of the block where no channels were observed was reported. A qualitatively similar morphology was observed in our experiments with predominantly clear ice at an inclination of \unit{30}{\degree}, where channels developed along the entire ice surface without the enhanced melting near the top that characterizes the opaque-ice experiments.

These channels form at low salinities, where the meltwater is denser than the ambient fluid and flows downward along the ice surface. Under these conditions, a Rayleigh--Bénard-type instability develops within the meltwater layer and is advected downward by the mean flow, giving rise to the observed channelized morphology, as proposed by Cohen et al. \cite{Cohen2020}. When salt is added, the density contrast between meltwater and ambient water changes. Above $R_\rho \approx 0.93$ (which is equivalent to a salinity concentration of 1.88 g/kg at \unit{19}{\celsius}), the meltwater becomes less dense than the surrounding fluid. Under these conditions, no convective instabilities exist and channels do not form. Experiments at \unit{30}{\degree} inclination confirm this transition; channels appear for $R_\rho$ below 0.93 but are absent at higher salinities. 

The difference between the morphologies observed for clear and opaque ice requires an additional explanation. The most notable distinction between the two ice types is their air content. Whereas the clear ice contains only a very small number of isolated bubbles, the opaque ice consists of a large population of small bubbles distributed throughout the ice. During melting, these bubbles are released and rise along the ice surface, in the opposite direction as the downward flow of meltwater. We further observed that the bubbles preferentially travel through the already-developed channels. Based on these observations, we hypothesize that the bubbles contribute to the subsequent evolution of the channelized morphology by locally enhancing melting. Such an effect would be expected to be strongest near the top of the ice face, where bubbles released over the entire ice surface converge as they rise.

Previous studies have demonstrated that entrapped air bubbles can modify melt dynamics. Wengrove et al. (2023) \cite{Wengrove2023} showed that \textit{pressurized} bubbles embedded within ice, such as those commonly present in icebergs and glacier, can increase melt rates by more than a factor of two relative to clear ice, resulting from an enhanced kinetic energy and turbulent kinetic energy within the boundary layer. In addition, bubbles have been shown to influence ice morphology as well. Their presence promotes surface roughening during melt, in contrast to the smooth melt surfaces observed for clear ice under similar environmental conditions \cite{Josberger1980}. 
Here, we hypothesize that entrapped air bubbles influence the development of both the surface morphology and the local melt rate of inclined ice blocks in the temperature-driven regime. However, a quantitative comparison with bubble-free ice was not possible because of the limitations of our imaging methodology. We emphasize that this proposed mechanism is specific to the conditions of the present experiments and should not be interpreted as an explanation for runnel formation in geophysical settings. Indeed, natural runnels are commonly observed under conditions of low temperature and elevated salinity, corresponding to large values of $R_\rho$, for which channel formation was not observed in the present study. 

In contrast, all experiments conducted at inclinations greater than $\theta > \unit{10}{\degree}$ exhibit a concave melt surface, with reduced melt rates along the lateral edges relative to the central region. We hypothesise that this pattern may be caused by edge effects. In the temperature-driven and competing regimes, the lateral thermal boundary layer descends along the sides, potentially thickening the local boundary layer and thereby reducing melt rates. This interpretation is likely less relevant in the salinity-driven regime, where stronger buoyancy forces are expected to drive predominantly upward flow.

For intermediate salinities,  $2 \lesssim R_\rho \lesssim 6$ (corresponding to about 6 g/kg and 25 g/kg respectively), the formation of scallops on the ice surface is observed (Figures \ref{fig:profile_salinity}$b$ and \ref{fig:profile_inclined}$b$), whereas a smooth surface is seen outside this range. Similar patterns have been observed in previous melting experiments \cite{Bushuk2019}, where scallops formed under forced, rather than natural, convection. The scalloped morphology region coincides with the competing flow regime. These results are consistent with Xu et al. \cite{xu2024}, where they also observe scallops for only intermediate salinities. Upon introducing an inclined surface, the $R_\rho$ range where scallops appear shifts. Notably, for higher angles, the critical $R_\rho$ at which scallops cease to form decreases. Section \ref{sec:scal_analysis} contains further analysis on the morphology of the scallops.

The variations of $\mathrm{Nu}$ with salinity (represented by $R_\rho$) and inclination is shown in Figure~\ref{fig:vertical_mr}. A clear non-monotonic dependence of the Nusselt number with salinity is seen, showing a minima around $R_\rho \approx 3$. Previous studies by Yang et. al. (2023) and Xu et. al. (2024) \cite{Yang2023,xu2024} show similar non-monotonic behavior. This reduction in heat transfer, and thus in melting rate, is explained by the competition between the saline-driven and thermal-driven buoyancy forces. Following the definition of $R_\rho$ we would expect the minima for $R_\rho = \mathcal{O}(1)$, given the non-linear density profile. Consistent with our prediction, a minimum is observed at $R_\rho \approx 3$. This trend persists for all tested inclination angles, although no systematic dependence on inclination is apparent (Figure~\ref{fig:vertical_mr}$b$). Earlier studies \cite{McConnochie2018,Mondal2019} reported a reduction in melting rate with increasing inclination, attributed to the weakening of buoyancy-driven flow along the ice surface. However, those papers focused on oceanic conditions, salinities near 35 g/kg and low ambient temperatures ($\unit{0}{\celsius} < T_\infty\leq \unit{5}{\celsius}$), resulting in $R_\rho > 50$, where meltwater primarily flows upward and a counterflow of cooled saline water is absent.

\subsection{Scallop analysis} \label{sec:scal_analysis}

\begin{figure}
    \centering
    \includegraphics[width=1.0\linewidth]{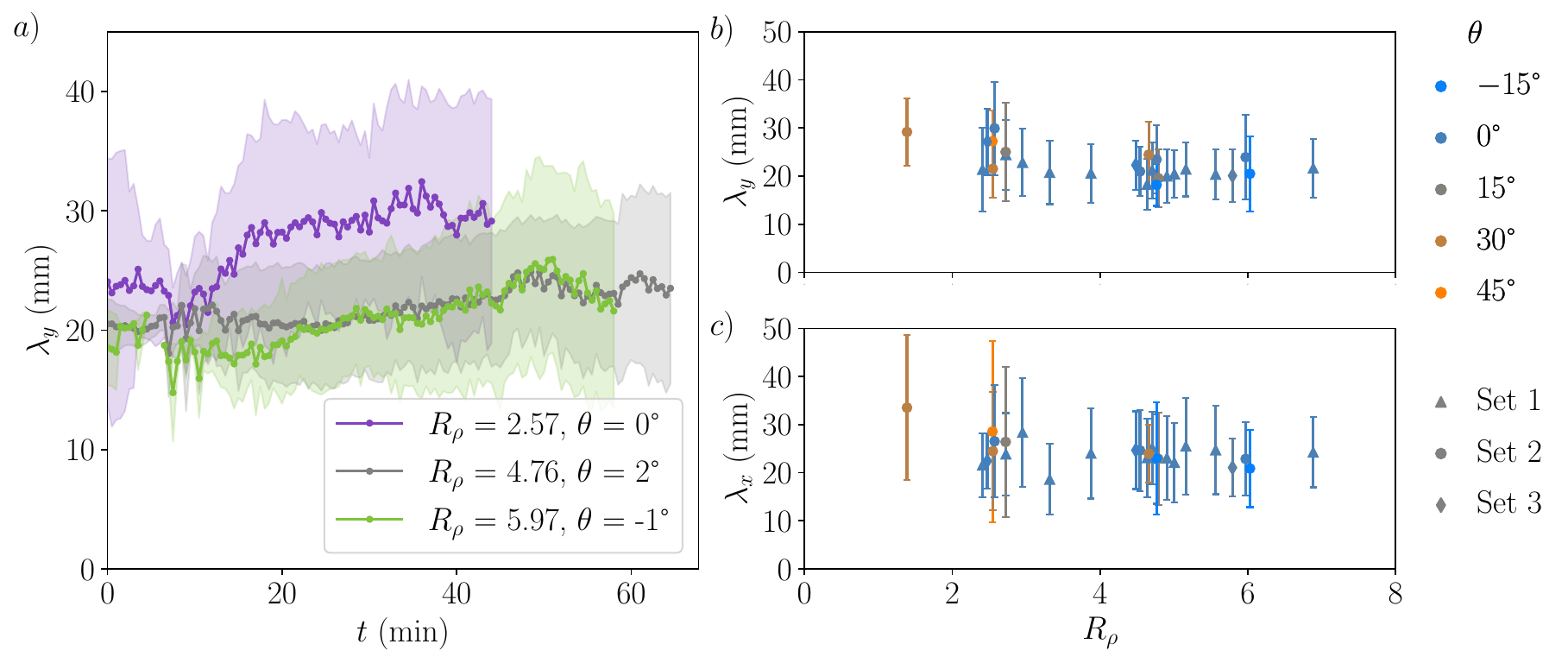}
    \caption{$a)$ Spatial mean of the vertical scallops wavelength development over time, with the spread between different scallops shown as shaded regions. The shown salinities are taken from Set 2 experiments and are representative of the behavior for all experiments that present scallops. $b)$ Spatial and temporal mean of the vertical scallop wavelength of the last 10 minutes of each experiment. The interval markers show the spread of the mean values over that time. $c)$ Analogous to $b)$ for the horizontal scallops wavelength.}
    \label{fig:walen}
\end{figure}

\begin{figure}
    \centering
    \includegraphics[width=1.0\linewidth]{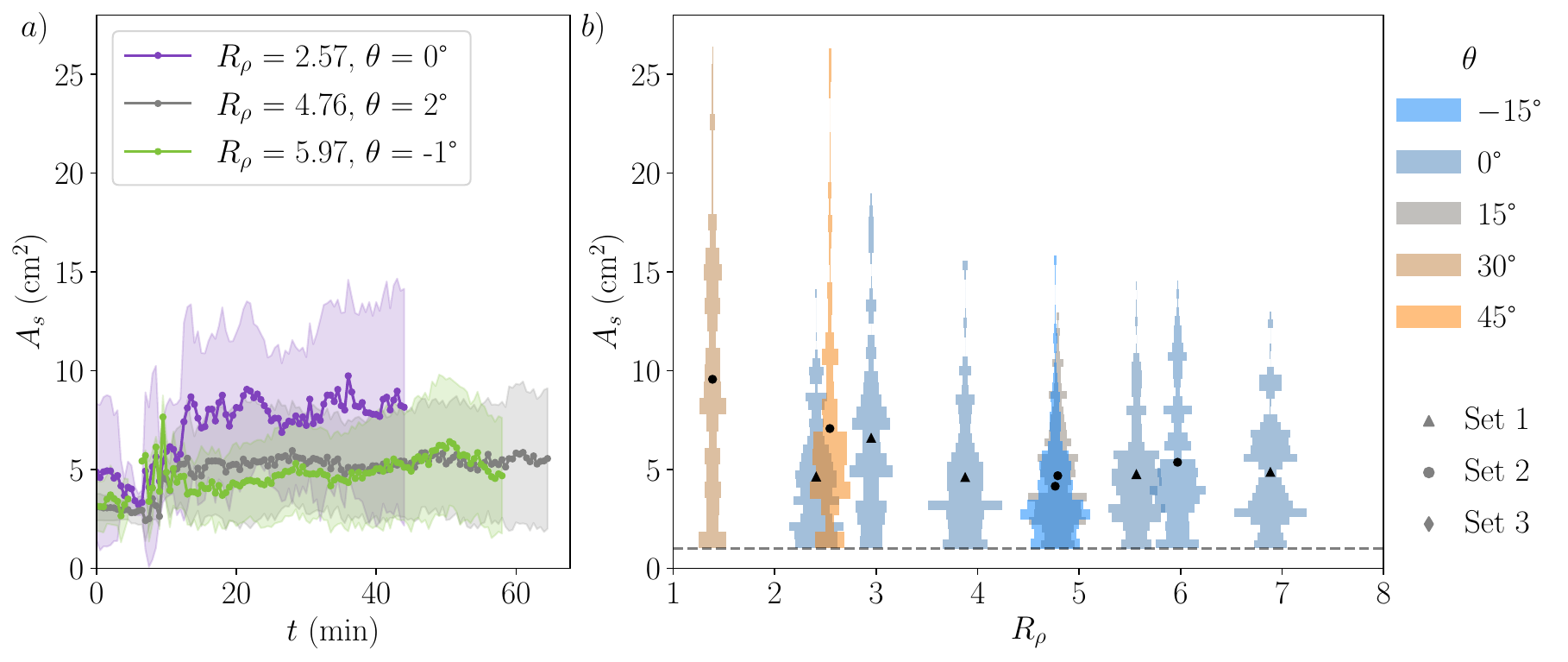}
    \caption{$a)$ Mean scallops area over time. The shown experiments are from Set 2. $b)$ Violin plot for the area distribution of scallops for the last 10 minutes of each experiment. The dashed line indicates the minimum cutoff area (1 cm$^2$) for scallops to be detected. }
    \label{fig:area}
\end{figure}

\begin{figure}
    \centering

    \includegraphics[width=1.0\linewidth]{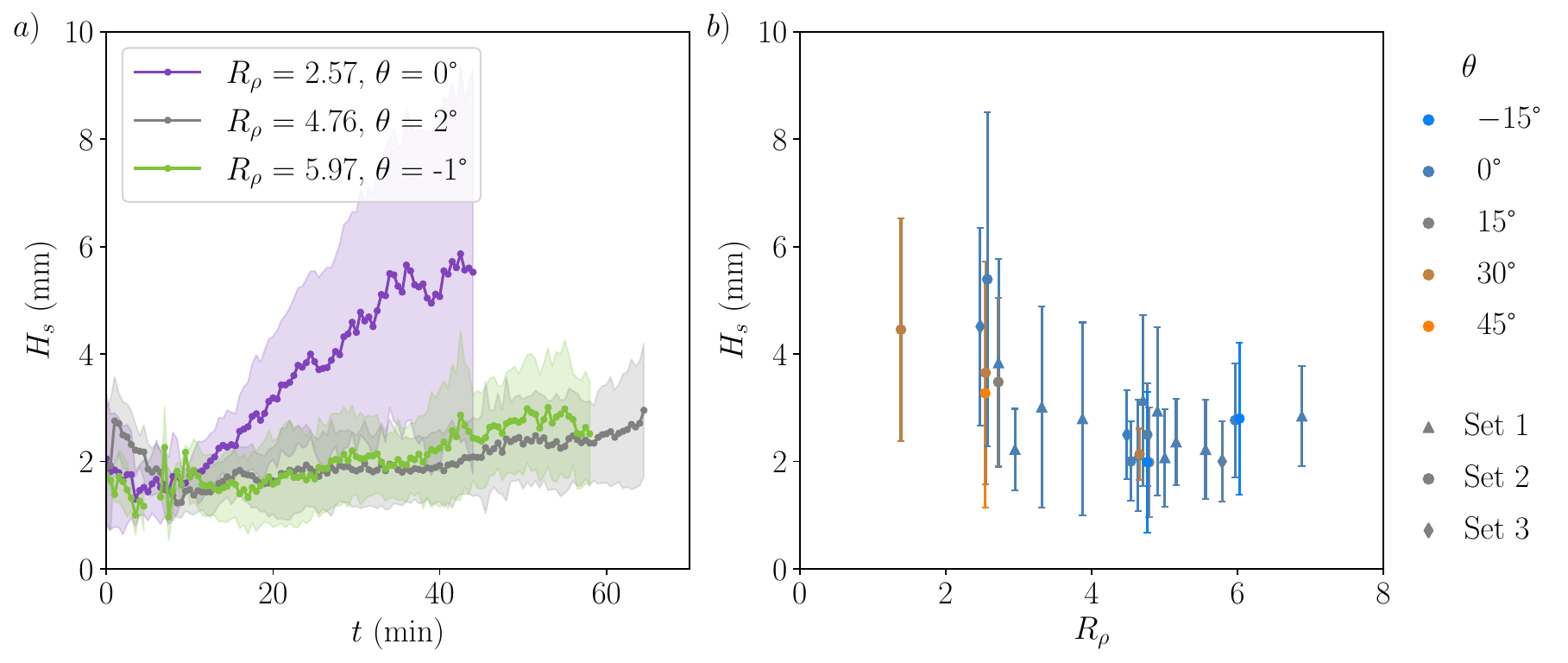}
    \caption{$a)$ Development of the amplitude $H_s$ of the scallops over time. The shown experiments are from Set 2. $b)$ Spatial and temporal mean of the amplitude over the last 10 minutes of each experiment. Interval markers show the spread of all scallops amplitudes over that time. }
    \label{fig:amplitude}
\end{figure}

To study the morphology of the scallops, a watershed algorithm was used to segment the surface of the ice into individual scallops. The watershed was applied to the curvature of the profile, rather than to its height, as this provided more consistent results. Following the watershed segmentation, not all regions segmented could be considered scallops. Regions that were too small (less than 1 cm$^2$) or too smooth were removed for the statistical analysis of the scallops properties. Appendix \ref{sec:watershed} contains further details on the watershed application and the exact criteria used for scallop detection.

With the scallops regions defined, various properties of the scallops can be calculated. We can define the typical horizontal and vertical sizes based on the moments of inertia
\begin{align}
    \lambda_x = \sqrt[4]{12} \,\, \sqrt[8]{\frac{I_y^3}{I_x}},& \quad \lambda_y = \sqrt[4]{12} \,\, \sqrt[8]{\frac{I_x^3}{I_y} } \\
    I_x = \iint_A (y - \langle y \rangle)^2 dA,& \quad I_y = \iint_A (x - \langle x \rangle)^2 dA
\end{align}
where $I_x$ and $I_y$ are the centralized second moments of area with respect to the $x$-axis and $y$-axis, respectively, and $A$ is the area of the scallop. This formulation is based on the centralized second moments of area for a rectangular shape, providing its base and height values. Any shape inscribed within that rectangle will yield a smaller value for $\lambda_x$ and $\lambda_y$ than the base and height of the rectangle. We found this definition to give robust estimations for the horizontal and vertical characteristic length scales of the scallops. Figure \ref{fig:walen}$a$ illustrates the mean vertical wavelength over time for three different salinities. The shaded regions depict the spread of wavelengths at each time point. Once development of scallops begins (typically between 10 and 15 minutes of starting the experiment), a slight increase in the mean vertical wavelength is observed. Figures \ref{fig:walen}$b$ and \ref{fig:walen}$c$ display the mean values during the final 10 minutes of the experiment. We choose to compare for the final 10 minutes since it corresponds to similar amount of recession of the ice front.  Notably, neither the vertical nor horizontal wavelengths exhibit significant variation across different salinities. Additionally, they are generally observed to be similar in value. However horizontal wavelengths show a much bigger spread, 50\% bigger on average. 

In the vertical direction, a viscous Kelvin--Helmholtz-type instability is often reported as cause for the scalloping \cite{xu2024,Weady2022}.
However, we do not observe the vertical wavelength scaling $\lambda_y \propto y^{-3/4}$ reported by Weady et al. \cite{Weady2022}. Claudin et al. \cite{Claudin2017} conducted a theoretical instability analysis based on a turbulent mixing framework that accounts for bed roughness, but their model includes only viscous and solutal boundary layers, which doesn't account for bidirectional flows like the ones present in our experiments. Moreover, a direct comparison with their predictions is not possible due to the absence of an estimate for the shear velocity in our experiments.
Despite these discrepancies, the presence of a vertical instability appears central to scallop formation \cite{xu2024} and would explain the low variability of the vertical wavelength compared to the horizontal direction. 
The emergence of three-dimensional patterns, instead of two-dimensional patterns perpendicular to the flow direction, remains as an open question.

Figure \ref{fig:area}$a$ shows the temporal evolution of scallop area for the same experiments presented in Figure \ref{fig:walen}$a$. Consistent with the behavior observed for the vertical wavelength, the scallop area undergoes a transition at approximately 10 minutes, coinciding with the start of scallop formation. After the scallops have developed, the mean area remains relatively stable, exhibiting fluctuations of only about $\pm 5\%$ around the mean. The distribution of scallop areas over the final 10 minutes of each experiment is shown in Figure \ref{fig:area}$b$. Both the spread of the distribution and the mean scallop area decrease with increasing $R_\rho$ up to approximately $R_\rho \approx 4$, beyond which no significant variation is observed. Thus, density ratios above $R_\rho \approx 4$ produce scallops that are smaller and more uniform than those observed at lower density ratios, but further increases in $R_\rho$ do not result in appreciable changes in their size or uniformity. It should be noted, however, that these distributions are computed from approximately 460 regions for the first data set and 617 for the second, and therefore may not be fully statistically converged.

To quantify the scallops amplitude, we define it as the difference between the minimum height within a scallop and the mean wall height that defines the scallop region. As shown in Figure \ref{fig:amplitude}$a$, scallop amplitudes continue to increase throughout the duration of the experiments. When averaged over the final 10 minutes, both the mean amplitude and the spread of the distribution decrease with increasing $R_\rho$ up to approximately $R_\rho \approx 4$ (Figure \ref{fig:amplitude}$b$), beyond which no significant variation is observed, consistent with the behavior of the scallop area.

\begin{figure}[t]
    \centering
    \includegraphics[width=1.0\linewidth]{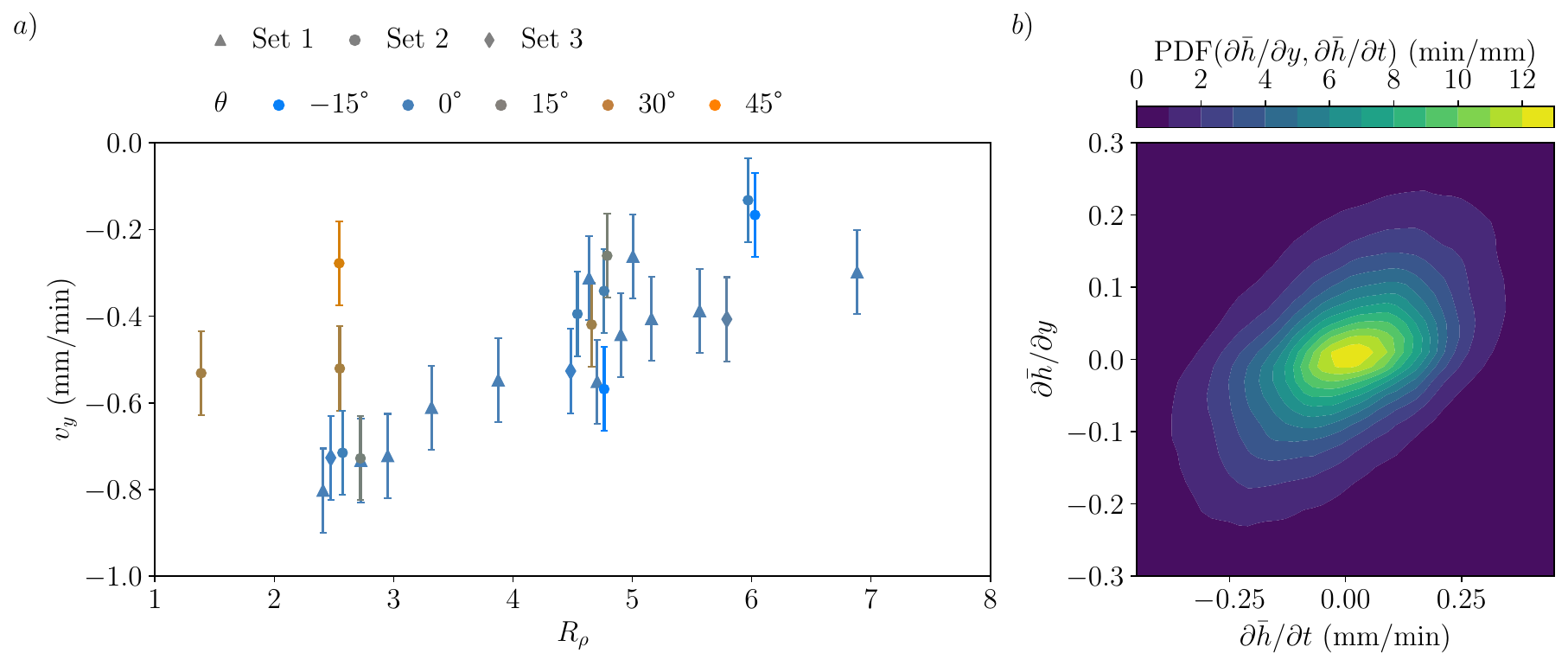}
    \caption{$a)$ Vertical velocity of the scallops. $b)$ Contour plot of the joint probability density function of the vertical and temporal derivatives of the residual height profile ($\bar{h}$) for the last 10 minutes of all scalloped experiments. Only regions identified as scallops were accounted for the joint probability density function. The Pearson correlation coefficient between $\partial \bar{h}/\partial t$ and $\partial \bar{h}/\partial y$ is 0.44.}
    \label{fig:vertical_vel}
\end{figure}

A prominent feature of the scallops is their downward migration over time. To quantify the migration velocity, we define the residual height profile
\begin{equation}
    \bar{h}(x,y,t) \equiv h(x,y,t) - P_4^t(x,y) 
\end{equation}
where $P_4^t$ denotes a fourth-order polynomial fit of $h(x,y)$ at time $t$. This detrending procedure isolates the scallop morphology and dynamics from the background profile. The positions of peaks in the residual height profile was tracked in time, and the migration velocity was obtained as the slope of a linear fit to the peak trajectories. The horizontal component of the velocity exhibits random fluctuations around zero, which is expected. Figure \ref{fig:vertical_vel}$a$ illustrates the variation in mean vertical velocity and its spread across the tracked peaks. As salinity increases, the mean vertical velocity decreases in magnitude, consistent with previous findings \cite{xu2024}. Figure \ref{fig:vertical_vel}$b$ shows a contour plot of the joint probability density function of $\partial \bar{h}/\partial t$ and $\partial \bar{h}/\partial y$. The probability density function was computed from data collected during the final 10 minutes of all experiments exhibiting scallops, considering only spatial regions identified as a scallop. 
A clear correlation between the temporal and vertical derivatives is observed, with the Pearson correlation coefficient being 0.44. This correlation indicates that melting rates ($-\partial \bar{h}/\partial t$) are higher at the top of a peak, where the vertical derivative is negative, compared with the bottom of a peak, where the vertical derivative is positive. This asymmetry in melting rates drives the downward migration of the scallops.

\section{Conclusions} \label{sec:conclusion}

Our experiments demonstrate that the melting morphology of inclined ice blocks is strongly governed by the relative contributions of thermal and saline buoyancy, as characterized by the density ratio $R_\rho$, and by the inclination angle $\theta$. We categorize the morphology of the melting ice blocks into fives categories: scalloped, channelized, top melting, bottom melting, and incurved. In order to obtain the morphologies, we implemented fringe projection profilometry (FPP) to measure the ice-water interface profile. A combination of ST-PSM and OSM was developed and applied to reduce errors from low contrast data and to eliminate non-physical stripe patterns appearing on the reconstructed surfaces.

In the temperature-driven regime ($R_\rho \lesssim 2$), two distinct morphologies emerge. At sufficiently large angles ($\theta \gtrsim \unit{15}{\degree}$), vertical channels form along the ice surface, originating from a Rayleigh–Bénard type instability. At lower inclinations, the flow regime promotes enhanced melting near the upper portion of the ice, producing a top-melting morphology.
 
In contrast, the salinity-driven regime ($R_\rho \gtrsim 6$) is characterized by preferential melting at the lower portion of the ice, in agreement with expectations from the flow regime. However, the transition point commonly associated with this regime, a cavity forming within the ice surface, was not observed. 

At the same time, all experiments with $\theta>\unit{10}{\degree}$  exhibited a concave (incurved) melt profile; however, this pattern is likely attributable to edge effects and interactions with the sidewalls, and therefore was not interpreted as a primary morphological regime. Lastly,  within the competing regime ($2 \lesssim R_\rho \lesssim 6$), we observe the formation of scallops on the melting surface. However, the range of $R_\rho$ over which scallops develop depends on the inclination angle.

For the channelized morphology, we observed that bubbles can influence the local melt rate of inclined ice blocks. By rising through previously made channels, the bubbles can locally enhance the melt rate, deepening them on the process. This effect is more clear at the top of the ice, where the channels were not expected to form. Previously, Josberger (1980) \cite{Josberger1980} demonstrated that bubbles can affect the morphology of ice, while Wengrove et al. (2023) \cite{Wengrove2023} showed that pressurized bubbles, which are present in natural glaciers, can increase the melt rate. It remains uncertain whether the bubbles present in our ice reduce the overall melt rate, or whether at higher salinities the bubbles—despite having no lasting effect on morphology—still affects the melt rate.

Analyzing the melt rate, we saw that it is not monotonic with salinity. We find a minima of the melt rate around $R_\rho \approx 3$, regardless of the angle. Previous studies show a similar behavior with salinity \cite{xu2024,Yang2023}. On the other hand, no clear trend appears with inclination. This is in contrast with the finding of a $\mathrm{Nu} \propto \cos(\theta)^{2/3}$ relation found in previous work \cite{McConnochie2018,Mondal2019}. This discrepancy is attributed to the different ambient conditions in our experiments, where down flows may occur (i.e. forced convection), unlike in the mentioned studies. 

Finally, using a watershed algorithm, we analyzed the morphology of the scalloped surfaces, namely the wavelength, area, amplitude, and migration velocity of the scallops. Both the vertical and horizontal wavelengths show no visible variation with $R_\rho$ or $\theta$.  However, vertical wavelengths exhibit less variability than horizontal wavelengths because they are set by an instability that develops primarily in the vertical direction. As $R_\rho$ increases smaller, shallower, and more uniform scallops develop on the surface of the ice, while at the same time their migration velocity decreases.  

\par\medskip
\noindent\textbf{Data availability statement.}
The data for the surface profiles that support the findings of this study is openly available at \cite{data}.
\par\medskip

\appendix
\section{Experimental conditions} \label{sec:table_exp}

\setlength{\tabcolsep}{10pt}
\renewcommand{\arraystretch}{1.2}

\Needspace{0.8\textheight}
\begin{longtable}{ccccccc}[t]
    \centering
        Set & $S_\infty$ (g/kg) & $\theta$ (\degree) & $T_\infty$ (\celsius) & Ra ($10^7$) & $R_\rho$ & Nu \\ 
        \hline \multirow{16}{*}{1}
        & 0.0 	& -2.0 	& 21.0 	& 0.41  & 0.00 	& 124 	\\
        & 2.3 	& 1.7 	& 20.6 	& 0.0054 & 0.92 & 107 	\\
        & 4.1 	& -0.6 	& 19.6 	& 0.34  & 1.74 	& 102 	\\
        & 6.4 	& 0.0 	& 20.1 	& 0.71  & 2.41 	& 107 	\\
        & 8.0 	& -0.5 	& 20.7 	& 0.94  & 2.73 	& 115 	\\
        & 8.0 	& 0.4 	& 19.9 	& 0.98  & 2.95 	& 90 	\\
        & 8.3 	& 1.1 	& 19.1 	& 1.1   & 3.32 	& 110 	\\
        & 10.3 	& 0.6 	& 19.2 	& 1.4   & 3.88 	& 109 	\\
        & 12.9 	& 1.3 	& 18.6 	& 1.9   & 4.90 	& 104 	\\
        & 14.8 	& -0.3 	& 20.0 	& 2.1   & 4.64 	& 99 	\\
        & 15.2 	& 1.1 	& 20.1 	& 2.2   & 4.71 	& 124 	\\
        & 15.2 	& 0.6 	& 19.5 	& 2.2   & 5.01 	& 102 	\\
        & 18.1 	& 1.8 	& 20.4 	& 2.6   & 5.16 	& 116 	\\
        & 21.9 	& 0.6 	& 18.6 	& 3.3   & 6.88 	& 130 	\\
        & 23.4 	& 2.1 	& 21.4 	& 3.4   & 5.56 	& 124 	\\
        & 27.4 	& 1.1 	& 17.8 	& 4.3   & 8.47 	& 138 	\\
        \hline \multirow{27}{*}{2}
        & 0.0 	& -17.5 & 19.3 	& 0.32  & 0.00 	& 131 	\\
        & 0.0 	& 0.3 	& 19.0 	& 0.32  & 0.00 	& 139 	\\
        & 0.0 	& 13.1 	& 19.3 	& 0.33  & 0.00 	& 142 	\\
        & 0.0 	& 14.9 	& 20.1 	& 0.36  & 0.00 	& 152 	\\
        & 0.0 	& 18.7 	& 19.4 	& 0.32  & 0.00 	& 156 	\\
        & 0.0 	& 34.5 	& 19.0 	& 0.26 	& 0.00 	& 171 	\\
        & 0.0 	& 43.7 	& 19.7 	& 0.26  & 0.00 	& 149 	\\
        & 1.2 	& 29.3 	& 19.3 	& 0.11  & 0.59 	& 141 	\\
        & 1.8 	& 29.1 	& 19.0 	& 0.013	& 0.89 	& 112 	\\
        & 2.9 	& 28.3 	& 19.0 	& 0.15 	& 1.39 	& 110 	\\
        & 6.8 	& 0.4 	& 20.0 	& 0.78  & 2.57 	& 131 	\\
        & 6.8 	& 15.6 	& 19.5 	& 0.77  & 2.72 	& 124 	\\
        & 6.8 	& 42.6 	& 20.1 	& 0.57  & 2.54 	& 104 	\\
        & 6.9 	& 44.7 	& 19.4 	& 0.59  & 2.79 	& 112 	\\
        & 7.0 	& -17.3 & 19.0 	& 0.82  & 2.95 	& 107 	\\
        & 7.0 	& 29.3 	& 20.3 	& 0.70  & 2.55 	& 116 	\\
        & 12.9 	& 0.4 	& 19.3 	& 1.8   & 4.54 	& 107 	\\
        & 13.0 	& -16.3 & 18.9 	& 1.8   & 4.76 	& 104 	\\
        & 13.2 	& 2.1 	& 19.0 	& 1.9   & 4.76 	& 102 	\\
        & 13.2 	& 47.7 	& 19.7 	& 1.2   & 4.43 	& 131 	\\
        & 13.3 	& 16.8 	& 19.0 	& 1.8   & 4.79 	& 143 	\\
        & 13.3 	& 28.8 	& 19.4 	& 1.7   & 4.59 	& 121 	\\
        & 19.7 	& -16.5 & 19.3 	& 2.8   & 6.03 	& 115 	\\
        & 19.8 	& 19.3 	& 19.8 	& 2.8   & 5.77 	& 158 	\\
        & 20.0 	& -0.9 	& 19.5 	& 3.0   & 5.97 	& 120 	\\
        & 20.1 	& 29.6 	& 19.2 	& 2.6   & 6.16 	& 160 	\\
        & 20.1 	& 31.2 	& 18.8 	& 2.6   & 6.41 	& 133 	\\
        \multirow{8}{*}{2}
        & 20.1 	& 50.7 	& 19.2 	& 1.9   & 6.16 	& 138 	\\
        & 26.1 	& -1.4 	& 19.8 	& 4.0   & 6.83 	& 143 	\\
        & 26.1 	& 15.7 	& 19.4 	& 3.8   & 7.08 	& 162 	\\
        & 26.2 	& 28.6 	& 19.1 	& 3.5   & 7.29 	& 157 	\\
        & 26.2 	& 47.0 	& 20.0 	& 2.7   & 6.73 	& 152 	\\
        & 34.3 	& -16.6 & 19.1 	& 5.1   & 8.44 	& 166 	\\
        & 34.5 	& -18.7 & 18.7 	& 5.1   & 8.76 	& 153 	\\
        & 34.5 	& 29.4 	& 19.2 	& 4.7   & 8.40 	& 142 	\\
        \hline \multirow{4}{*}{3}
        & 5.8 	& -1.7 	& 19.2 	& 0.24  & 2.47 	& 90 	\\
        & 12.3 	& -2.7 	& 19.1 	& 0.64  & 4.48 	& 86 	\\
        & 12.4 	& 27.1 	& 18.8 	& 0.58  & 4.66 	& 104 	\\
        & 18.5 	& 5.0 	& 19.3 	& 1.0   & 5.79 	& 114 	\\
    \caption{ Experimental conditions. The ambient salinity and temperature of the ambient water temperatures $S_\infty$ and $T_\infty$, the inclination of the ice block $\theta$, Rayleigh number Ra, density ratio $R_\rho$, and Nusselt number Nu.} \\
    \label{tab:experiments}
\end{longtable}

Table \ref{tab:experiments} summarizes the experimental conditions for all experiments conducted in this study. The datasets are grouped into Sets 1, 2, and 3 according to ice-block dimensions and the phase-shifting profilometry (PSP) method used for surface reconstruction. Set 1 corresponds to ice blocks with dimensions $H \times W \times L = 32~\mathrm{cm} \times 17~\mathrm{cm} \times 10~\mathrm{cm}$, for which surface profiles were obtained using ST-PSM. Sets 2 and 3 employed OST-PSM for surface reconstruction. Ice blocks in Set 2 had dimensions $32~\mathrm{cm} \times 23~\mathrm{cm} \times 12~\mathrm{cm}$, while those in Set 3 measured $23~\mathrm{cm} \times 15~\mathrm{cm} \times 12~\mathrm{cm}$. Additionally, set 1 initial ice temperature is \unit{-16}{\celsius}, while for sets 2 and 3 is \unit{-5}{\celsius}.

\section{Scallop segmentation} \label{sec:watershed}

Scallops were identified using a watershed segmentation algorithm applied to the local curvature field of the ice surface. The local curvature was computed as
\begin{equation}
    k = \frac{(1 + \bar{h}_y^2) \bar{h}_{xx} + (1 + \bar{h}_x^2) \bar{h}_{yy} - 2 \bar{h}_x \bar{h}_y \bar{h}_{xy}}{(1 + \bar{h}_x^2 + \bar{h}_y^2)^{3/2}},
\end{equation}
where the subscripts indicate the derivative with respect to that variable. Prior to differentiation, a Gaussian filter with a standard deviation of 6 mm was applied to the height profile $h$ to reduce noise. Because the watershed was applied to $-k$, with the aim of identifying scallops as minima in $h$, we further suppressed the influence of positive curvature values. To do so, an additional Gaussian filter was applied to the positive portion of $k$ (by setting values of $k<0$ to zero), using standard deviations of 8 mm and 17 mm in the $x$ and $y$ directions, respectively.

\begin{figure}
    \centering
    \includegraphics[width=0.99\linewidth]{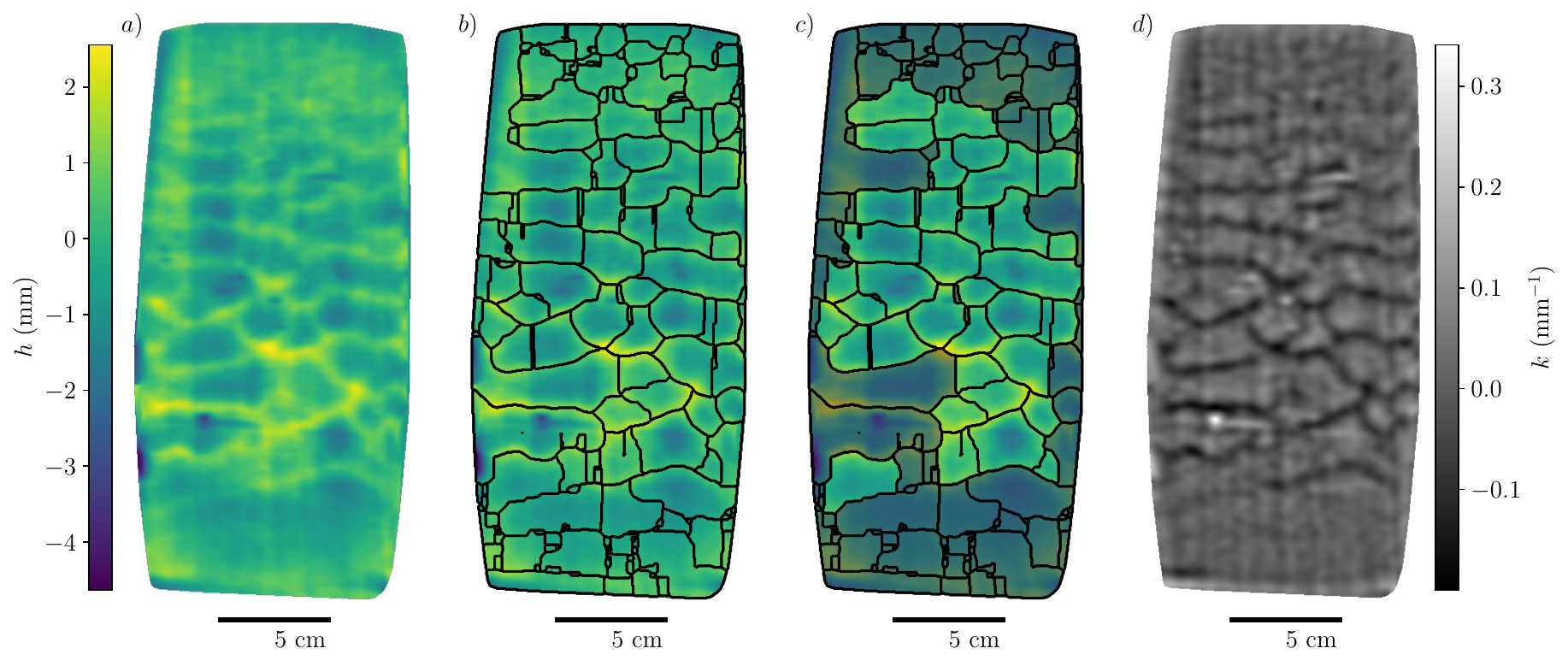}
    \caption{$a)$ Height profile of the ice surface, subtracted with a fourth order polynomial fit of the experiment $S=14.8$ g/kg, $T_w = \unit{20.0}{\celsius}$, $\theta = \unit{-0.3}{\degree}$, Ra $= 2.1 \times 10^7$, and $R_\rho=4.6$, at 30 minutes after the start of the experiment. The polynomial subtraction was performed in order to remove the height dependent background melt rate, improving visibility of the scallops. $b)$ Segmentation of the ice block using the watershed algorithm. $c)$ Greyed out areas correspond to regions that are not considered scallops with our filter. $d)$ Curvature of the height profile.  }
    \label{fig:watershed}
\end{figure}

Not all regions produced by the watershed correspond to true scallops. Regions with areas smaller than $1$ cm$^2$ were removed, as they were typically associated with noise and were disproportionately represented in the area distribution. Regions exhibiting a standard deviation of the curvature below 0.025 mm$^{-1}$ were also excluded because they were deemed too flat to represent scallops. Finally, regions reaching the boundaries of the ice block were taken out as well to avoid accounting for edge effects. Figure \ref{fig:watershed} illustrates the full scallop detection procedure for the experiment shown in Figure \ref{fig:profile_salinity}$b$.

\section{Morphology classification} \label{sec:morphologies}

\begin{figure}
    \centering
    \includegraphics[width=1.\linewidth]{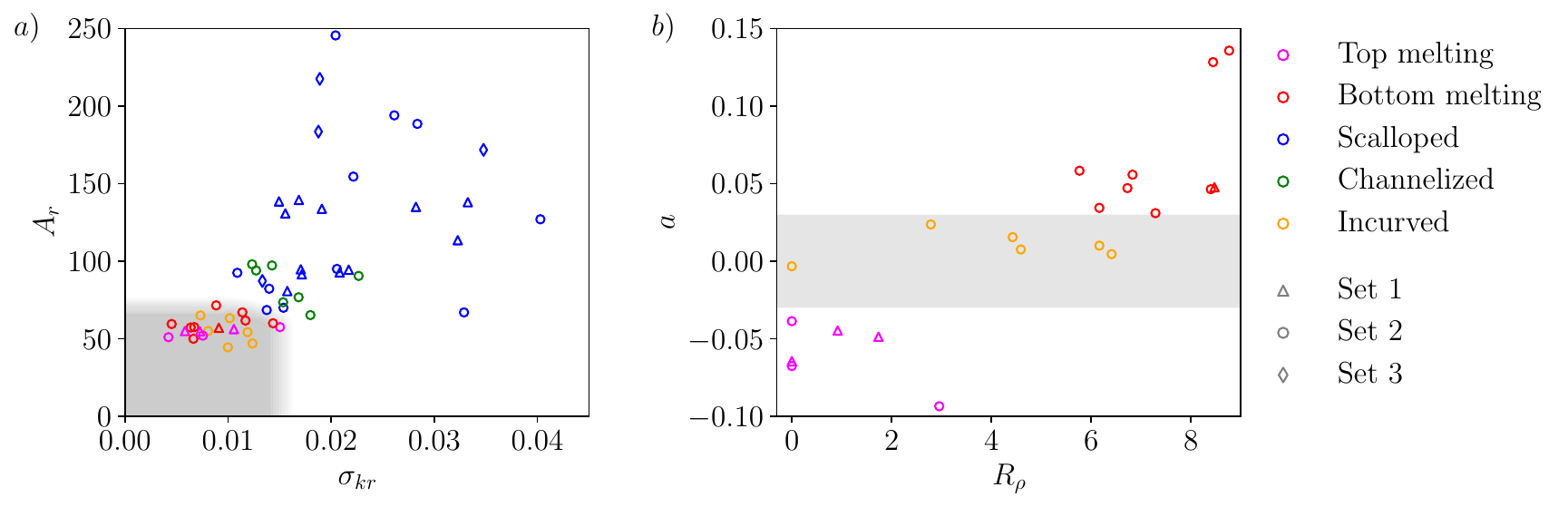}
    \caption{$a)$ Average region area $A_r$ against the average region curvature standard deviation $\sigma_{kr}$ for all experiments. The gray area indicates the region of the $A_r-\sigma_{kr}$ space within which surface profiles are classified as smooth. $b)$ Mean value of the fitted vertical slope coefficient $a$ for the remaining (smooth) surface profiles. The gray area indicates the threshold region $|a|\leq 0.03$ used to demarcate the smooth-surface morphologies.}
    \label{fig:separation}
\end{figure}

The classification of surface morphologies for each experiment was performed in two stages. First, we identified rough morphologies (scalloped or channelized), and subsequently distinguished among smooth surface morphologies (top melting, bottom melting, and incurved).

We start by applying the watershed algorithm described in Appendix \ref{sec:watershed} to segment each surface profile into discrete regions. For each experiment, we then computed the mean region area $A_r$ and the standard deviation of the region curvature $\sigma_{kr}$, averaged over the final 10 minutes of the experiment. It is important to note that, unlike in Appendix \ref{sec:watershed}, no removal of segments was applied; all regions produced by the watershed were included in the averaging. Figure \ref{fig:separation}$a$ shows $A_r$ versus $\sigma_{kr}$ for all experiments. Smooth morphologies are observed around the region $A_r \lesssim 70$ and $\sigma_{kr} \lesssim 0.015$, corresponding to the gray region in Figure \ref{fig:separation}$a$. Because these are not hard boundaries, the borders of this region are diffuse. Experiments lying near this boundary were visually inspected to determine whether surface roughness was present. Those experiments exhibiting roughness were classified as either scalloped or channelized; channelized cases were then separated from scalloped ones by visually identifying the presence of distinct vertical channels.

For the remaining experiments, which did not exhibit rough surfaces, the height profiles were fit with a plane of the form $h = ay+bx+c$. The coefficient $a$ was used to distinguish top melting from bottom melting morphologies. Figure \ref{fig:separation}$b$ shows the average value of $a$ over the final 10 minutes for all such experiments. A threshold of $\pm0.03$ was chosen to differentiate among morphologies. Experiments satisfying $-0.03\leq a \leq 0.03$ all possessed inclinations $\theta>\unit{10}{\degree}$, which matched the subset exhibiting an incurved surface profile. For this reason, this category was designated as incurved.

\bibliography{ms}

\end{document}